\documentclass[apj]{emulateapj}

\usepackage{apjfonts,ifthen,graphicx,times,amsmath,ifpdf}

\newcommand{\swift}{\textit{Swift}}
\newcommand{\swiftbat}{\textit{Swift}-BAT}
\newcommand{\swiftxrt}{\textit{Swift}-XRT}
\newcommand{\asca}{ASCA}

\newcommand{\xmm}{\textit{XMM-Newton}}
\newcommand{\chandra}{\textit{Chandra}}
\newcommand{\suzaku}{\textit{Suzaku}}
\newcommand{\integral}{INTEGRAL}

\newcommand{\degree}{\ensuremath{^{\circ}}}

\begin{document}

\title{The 70 Month \swiftbat\ All-Sky Hard X-Ray Survey}

\shorttitle{SWIFT-BAT 70 MONTH HARD X-RAY SURVEY}
\shortauthors{BAUMGARTNER ET AL}

 \journalinfo{(submitted to the Astrophysical Journal Supplement
   Series 16 Nov 2012)}

\author{
W.~H.~Baumgartner\altaffilmark{1,2,4},
J.~Tueller\altaffilmark{1},
C.~B.~Markwardt\altaffilmark{1},
G.~K.~Skinner\altaffilmark{1,3,4,6},\\
S.~Barthelmy\altaffilmark{1},
R.~F.~Mushotzky\altaffilmark{3},
P.~Evans\altaffilmark{5},
N.~Gehrels\altaffilmark{1}
}

\altaffiltext{1}{NASA/Goddard Space Flight Center, Astrophysics
 Science Division, Greenbelt, MD 20771}
\altaffiltext{2}{Joint Center for Astrophysics, University of
 Maryland Baltimore County, Baltimore, MD 21250}
\altaffiltext{3}{Department of Astronomy, University of
 Maryland College Park, College Park, MD 20742}
\altaffiltext{4}{CRESST/ Center for Research and Exploration in Space
  Science and Technology, 10211 Wincopin Circle, Suite 500, Columbia,
  MD 21044}
\altaffiltext{5}{X-Ray and Observational Astronomy Group/ Department
  of Physics and Astronomy, University of Leicester, Leicester, LE1
  7RH, United Kingdom}
\altaffiltext{6}{Max-Planck Institut f\"ur extraterrestriche Physik, 85748 Garching, Germany}
\altaffiltext{7}{Corresponding author: whbaumga@alum.mit.edu}

\begin{abstract}

We present the catalog of sources detected in 70 months of
observations of the BAT hard X-ray detector on the \swift\ gamma-ray
burst observatory.  The \swiftbat\ 70 month survey has detected 1171
hard X-ray sources (more than twice as many sources as the previous
22~month survey) in the 14--195~keV band down to a significance level
of 4.8$\sigma$, associated with 1210 counterparts.  The 70~month
\swiftbat\ survey is the most sensitive and uniform hard X-ray all-sky
survey and reaches a flux level of 1.03$\times
10^{-11}$~ergs~sec$^{-1}$~cm$^{-2}$ over 50\% of the sky and
1.34$\times 10^{-11}$~ergs~sec$^{-1}$~cm$^{-2}$ over 90\% of the sky.
The majority of new sources in the 70~month survey continue to be AGN,
with over 700 in the 70~month survey catalog.

As part of this new edition of the \swiftbat\ catalog, we also make
available 8-channel spectra and monthly-sampled lightcurves for each
object detected in the survey at the \swiftbat\ 70~month 
website.\footnote{\scriptsize \texttt{http://swift.gsfc.nasa.gov/docs/swift/results/bs70mon/}}

\end{abstract}




\section{Introduction}

The \swift\ Gamma-ray burst observatory \citep{gehrels04} was launched
in November 2004, and has been continually observing the hard X-ray
(14--195~keV) sky ever since with the Burst Alert Telescope (BAT).  We
have previously published BAT survey catalogs covering the first three
months of data \citep{markwardt05}, AGN detected in the first 9~months
of data \citep{tueller9}, and a complete catalog of sources detected
in the first 22 months of data \citep{tueller22}.  This paper extends
this work to include all sources detected in the first 70 months of
data between December~2004 and September~2010.

The main advances of the BAT 70~month survey compared to previous
\swiftbat\ surveys include better sensitivity resulting from a
complete reprocessing of the data with an improved data reduction
pipeline, the publication of 8 channel spectra, lightcurves sampled every
month throughout the mission, and a lower flux threshold resulting
from nearly a factor of three more integration time.

Hard X-ray source catalogs have also been published based on
observations from the \integral\ satellite \citep{4thbird,krivonos10}
and on independent analyses of \swiftbat\ data by the Palermo group
\citep{palermo39,palermo54} and by \cite{marco60,burlon11,voss10}.
The \integral-based surveys benefit from the somewhat better angular
resolution of the IBIS instrument ($\sim 12$~arcmin versus 19.5~arcmin
FWHM for BAT); however, the much narrower field of view of IBIS
coupled with their observing strategy limits the uniformity of the
IBIS sky coverage. This leads to lower sensitivity than BAT surveys
over much of the sky away from the galactic plane.  In the BAT surveys
the exposure is not biased towards the galactic plane, which leads to
deeper and more uniform coverage of evenly distributed objects like
AGN.

The \swiftbat\ 70~month survey has paid special attention to compiling
a uniform catalog by using a well-defined significance threshold and
energy band for inclusion of sources into the catalog. Particular
attention has been paid to the identification of sources, for which
examination of 3--10~keV X-ray data is crucial.  The BAT survey
catalogs of \cite{palermo39} and \cite{palermo54} often base their
counterpart identification on nearby ROSAT sources; \cite{tueller22}
have shown that the soft X-ray (0.1--2~keV) ROSAT fluxes are not well
correlated with BAT fluxes and could lead to incorrect counterpart
associations, especially in the galactic plane.  The BAT catalogs of
\cite{marco60,burlon11,voss10} often use the counterpart associations
of \cite{palermo54}, and utilize only part (15--55 keV) of the full
BAT energy band (14--195 keV).

Section~\ref{procedure} describes the data reduction and analysis
techniques used in the 70~month BAT survey, concentrating on the
improvements made in the data reduction pipeline.
Section~\ref{catalog} presents the 70~month catalog of detected
sources, spectra, light curves, and a characterization of the
properties of the survey.

\section{The \swift\ Mission and the BAT Instrument}
\label{bat_instrument}

The Burst Alert Telescope (BAT) on the \swift\ gamma-ray burst
observatory is a large coded-mask telescope optimized to detect
transient GRBs and is designed with a very wide field of view of $\sim
60 \times 100$~degrees.  \swift's general observation strategy is to
observe pre-planned targets with the narrow field of view X-ray
telescope (XRT, which is co-aligned with BAT) until a new GRB is
discovered by BAT, at which time \swift\ automatically slews to the
new GRB to follow up with the narrow field instruments for a couple of
days until the X-ray afterglow is below the XRT detection limit.  This
observation strategy is also very well suited for conducting an all sky
survey with BAT.

The \swiftbat\ survey's most important feature in comparison to the
similar \integral\ survey \citep{4thbird, krivonos10} is its uniform
sky coverage.  The very wide instantaneous field of view ($\sim$ 1/6
of the entire sky at 5\% partial coding) allows coverage of a
relatively very large fraction of the sky with each pointing.  The
effectively random pointing plan caused by GRB observations and
followup enables coverage of all parts of the sky.  The combination of
these two properties results in very uniform coverage of the entire
sky.

\begin{figure*}
  \begin{center}
    \ifpdf
      \resizebox{!}{1.72in}{\includegraphics{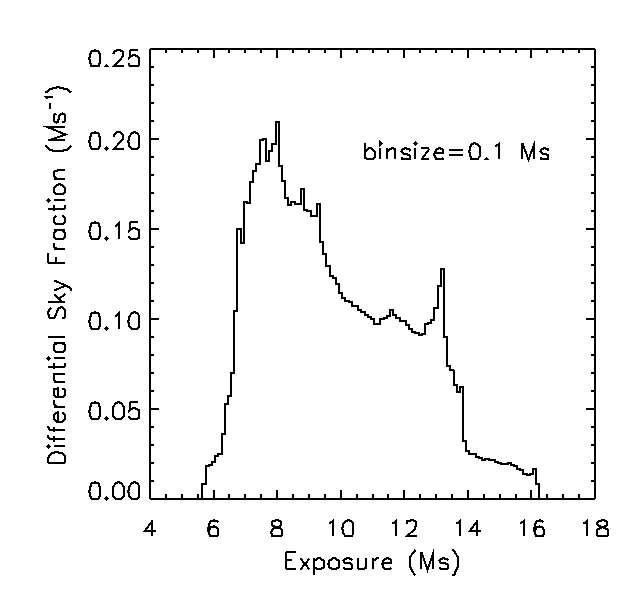}}
      \resizebox{!}{1.72in}{\includegraphics{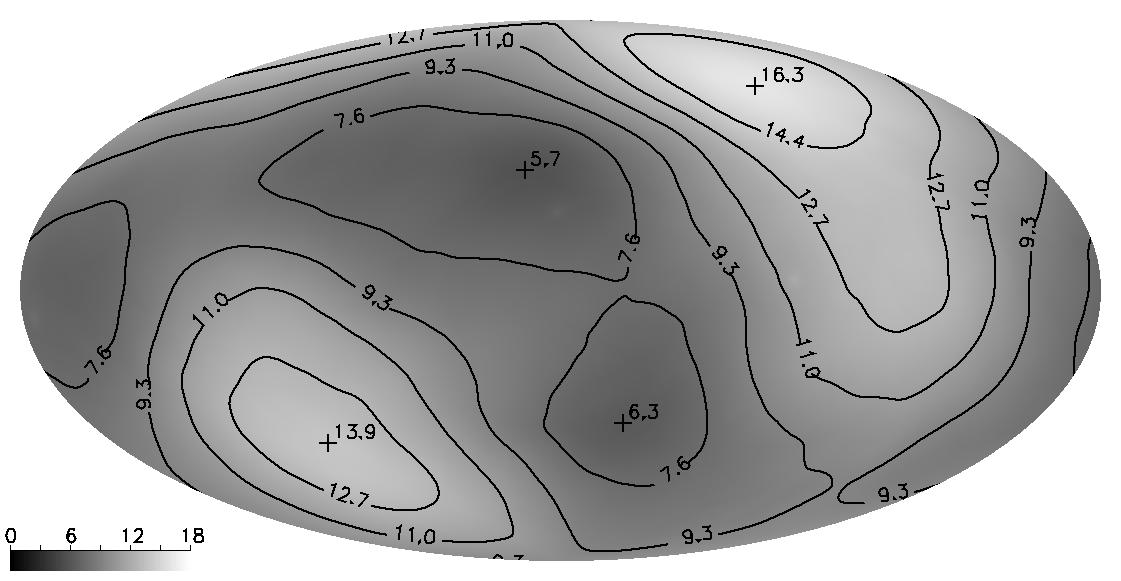}}
      \resizebox{!}{1.72in}{\includegraphics{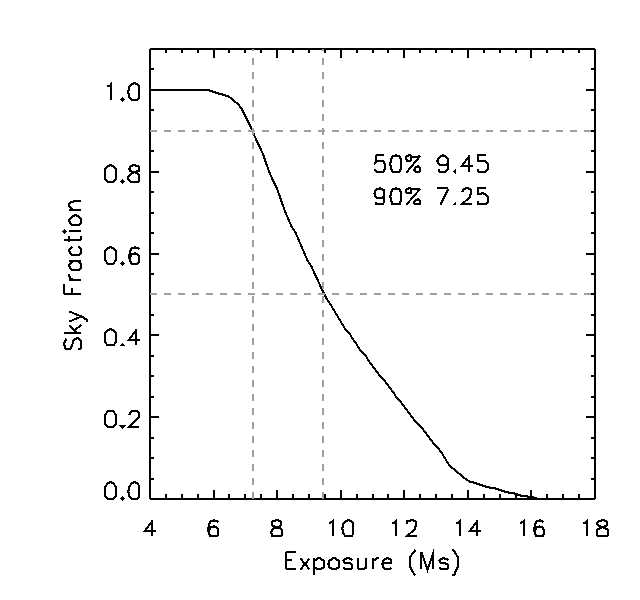}}
    \else
      \resizebox{!}{1.72in}{\includegraphics{expo_hist.eps}}
      \resizebox{!}{1.72in}{\includegraphics{expo_aitoff_2.eps}}
      \resizebox{!}{1.72in}{\includegraphics{expo_int.eps}}
    \fi
  \end{center}
\caption{Exposure in the 70~month \swiftbat\ survey.  The left panel
  shows the distribution of observations times across the sky, the
  center panel shows an all-sky exposure map in a galactic projection
  (with greyscale indicating megaseconds of exposure), and the right
  panel shows the fraction of sky covered as a function of exposure
  time). The areas of highest exposure in the all sky map are near the
  ecliptic poles because of \swift's sun and moon avoidance
  constraints.\label{exposure_fig}}
\end{figure*}
Figure~\ref{exposure_fig} illustrates the sky coverage in the 70~month
survey.  The center panel is an all-sky exposure map in galactic
coordinates, with the ecliptic plane marked.  \swift's pointing
constraints (primarily the sun and moon constraints) lead to areas of
highest exposure at the ecliptic poles.  The first panel in
Figure~\ref{exposure_fig} shows the distribution of exposure times in
the survey, and the third panel shows the fraction of the sky covered
as a function of exposure time.  Over 50\% of the sky has been
observed by \swiftbat\ for over 9.45~Ms in the 70~month survey, while
90\% of the sky is covered at the 7.25~Ms level.

Table~\ref{bat_info} provides some instrument parameters for
\swiftbat.  A full description of the BAT instrument can be found in
\cite{barthelmy05}.

\begin{deluxetable}{lll}[h]
\tablecaption{\swiftbat\ Instrument Parameters \label{bat_info}}
\tablewidth{\columnwidth}
\tablehead{
  \colhead{Parameter} & \colhead{Value} & \colhead{Comment}
}
\startdata
Energy Range & 14--195\,keV & \\
Field of View & $\sim85\degree \times \sim120\degree$ & 0\% partial coding \\
              & 2.29~sr & \ 5\% partial coding \\
              & 1.18~sr & 50\% partial coding \\
              & 0.34~sr & 95\% partial coding \\
Point Spread Function & 19.5\arcmin & Mosaicked sky maps \\
                      & 22\arcmin & in center of snapshot FOV \\
                      & 14\arcmin & at corners of snapshot FOV \\
Detector Area & 5243\,cm$^2$ & 32,768 CdZnTe detectors,\\
 & & 4\,mm $\times$ 4\,mm $\times$ 2\,mm \\
Aperture & 50\% open & Coded mask, random pattern \\
Coded Mask    & $\sim52,000$ tiles & 5\,mm $\times$ 5\,mm $\times$ 1\,mm Pb\\
Pointing constraints & $> 45\degree$ & Sun \\
                     & $> 30\degree$ & Earth limb \\
                     & $> 20\degree$ & Moon 
\enddata
\end{deluxetable}

\section{Procedure}
\label{procedure}

The data reduction and analysis for the BAT 70~month survey are based
on the procedures used in the BAT 22~month survey.  The complete
analysis pipeline is described in the BAT 22 month paper,
\cite{tueller22}.  

The data analysis and catalog generation process can be briefly
summarized as follows: The data from each snapshot (a single
\swift\ pointing of $\sim 20$~minutes) are extracted in the eight
bands listed in Table~\ref{energy_bands} and combined into all-sky
mosaic images.  The eight band mosaics are combined into a total band
map and a blind search for sources is done by finding all pixels above
the 4.8$\sigma$ detection threshold that are higher than each of their
immediately surrounding neighbors.  Then we use the rough positions of
these blind sources as inputs to a stage where we more carefully fit
for the BAT positions of these sources.  After that, we identify
counterparts to the blind sources by searching archival X-ray images
from high resolution instruments like \swiftxrt, \chandra, and \xmm.
This step allows the identification of previously-known X-ray sources
as well as uncataloged X-ray point sources that can be checked against
databases such as SIMBAD and NED for associations with objects in
other wavebands (such as galaxies in the 2MASS extended source
catalog).

Finally we associate the blind sources with counterparts by searching
the counterparts list for objects within a fixed match radius of each
blind source.

\begin{deluxetable}{cccrrrr}
\tablecaption{Energy Bands in the \swiftbat\ 70~Month Survey\label{energy_bands}}
\tablewidth{0pt}
\tablehead{
\colhead{Band} & \colhead{Low} & \colhead{High}&
\colhead{Crab}& \colhead{Rate}& \colhead{Crab} & \colhead{Crab}\\
\colhead{} & \colhead{[keV]} & \colhead{[keV]}& \colhead{Rate\tablenotemark{a}}& \colhead{Error\tablenotemark{b}}& \colhead{Flux\tablenotemark{c}}& \colhead{Weights\tablenotemark{d}}
}
\startdata
1       & 14  & 20  & 101.7 & 2.0 & 3.81 & 27.000  \\
2       & 20  & 24  & 57.0  & 1.3 & 1.87 & 35.260 \\      
3       & 24  & 35  & 100.0 & 2.2 & 3.71 & 22.700  \\
4       & 35  & 50  & 60.1  & 1.5 & 3.32 & 29.444 \\
5       & 50  & 75  & 48.7  & 1.5 & 3.56 & 21.272 \\
6       & 75  & 100 & 17.9  & 1.1 & 2.40 & 16.062 \\
7       & 100 & 150 & 9.3   & 1.0 & 3.21 & 8.449  \\
8       & 150 & 195 & 1.4   & 0.7 & 1.98 & 2.630
\enddata
\tablenotetext{a}{[$10^{-4}$~cts\,s$^{-1}$\,detector$^{-1}$].}
\tablenotetext{b}{[$10^{-6}$~cts\,s$^{-1}$\,detector$^{-1}$].  Computed from the rms
noise in a large annulus around the Crab (see~\S\ref{flux}).}
\tablenotetext{c}{Calculated from Equation~\ref{eqn:crab_cts}, in units of [$10^{-9}$ergs\,s$^{-1}$cm$^{-2}]$.}
\tablenotetext{d}{These weights are used when combining the 8
individual band maps into the total band map for source detection
purposes.  See \S~\ref{crabweighting}.}
\end{deluxetable}

\subsection{Data Processing Improvements}

We have kept the data reduction procedures in the 70~month survey the
same as in the 22~month survey except for improvements in these areas:
a gain correction for each of the individual CZT detectors, a more
thorough cleaning of bright sources from the snapshot images, and a
finer pixelization in the mosaicked all-sky maps.  

These changes have necessitated a complete reprocessing of all the
data from the entire BAT survey to ensure uniformity and homogeneity.
This complete reprocessing is different from our past procedure of
processing new data periodically (on a several month time scale) and
incorporating improvements to the data reduction pipeline before each
processing run of new data.  The current complete reprocessing of all
the data for the 70~month survey should lead to more homogeneous
results and a more uniform survey than the incremental processing used
for preparing previous catalogs.

We have also introduced minor changes in the data analysis and catalog
generation procedures.  These include the handling of confused
sources, the fitting of spectra and the calculation of the flux, the
match radius for finding counterparts, computing the spectral index
instead of hardness ratio, and the use of new Crab-derived
weights for combining energy bands for source detection.  These
changes are described in Sections~\ref{catalog} and \ref{crabweighting}.

\subsection{Gain Correction}

\begin{figure}
  \begin{center}
    \ifpdf
      \resizebox{\columnwidth}{!}{\includegraphics{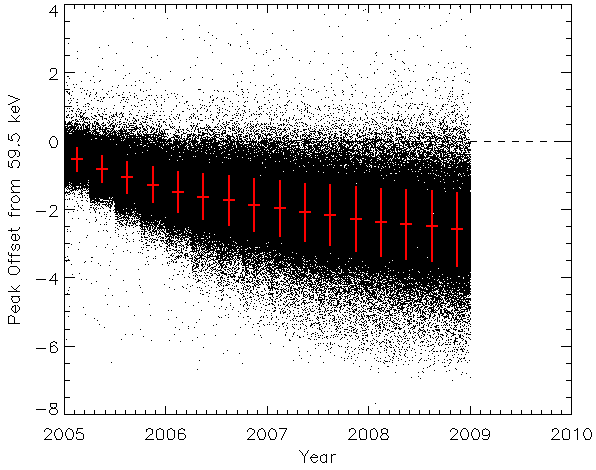}}
    \else
      \resizebox{\columnwidth}{!}{\includegraphics{respub_fullresid_2.eps}}
    \fi
  \end{center}
\caption{BAT detector gain as function of time through out the first
  58~months of the \swift\ mission.  The $y$-axis is the offset (in
  keV) between the location of the measured and nominal 59.5~keV line
  from the $^{241}$Am calibration source.  The $x$-axis is broken down
  into successive three month periods in which a black point is
  plotted for each of the 32,768 detector elements in the BAT array.
  The red crosses are the 3 month averages (and rms spread) for all
  the detectors in the entire BAT array.
\label{gainplot}}
\end{figure}

Analysis of BAT calibration spectra over the course of the
\swift\ mission shows that the detector energy channel containing the
59.5~keV peak from the $^{241}$Am tagged source has gradually shifted
as a function of time (see Figure~\ref{gainplot}).
This shift of up to about 3\% in the BAT energy scale is due to a
continuing small decrease in the gains of the individual CZT
detectors.  The data processing pipeline used in the analysis of
9~month and 22~month versions of the \swiftbat\ survey catalog did not
take this gain variation into account, leading to slightly low
energies being assigned to the detected X-rays.

For most detectors, this lead to a small change in the fluxes detected
in each of the 8 spectral channels---individual photons had too low of
an energy assigned to them, and counts shifted to lower channels or
below the low energy threshold.  Detector to detector variations in
gain also cause imaging noise as counts from different energy bands
fall into and out of the expected mask shadow in the detector plane.

In order to correct for the gain decreases in BAT, we have fit the
position of the peak channel of the 59.5~keV calibration line as a
function of time throughout the mission.  We then use the peak channel
position to determine a gain correction factor for each detector as a
function of time.  This gain correction factor is then used to correct
the bin edge location in channel space of the BAT energy
bands.

The physical origins of the gain shifts in CZT are not fully known.  A
leading hypothesis is that over time radiation damage creates more
charge traps in the CZT, reducing the charge collection efficiency and
hence reducing the gain.  There is also some indication that pixels
with the best charge collection properties at launch suffered the
largest gain decreases over time, while pixels with lower charge
collection efficiencies (and $\mu\tau$ products) suffer less from gain
decreases.  The differing susceptibility of each pixel to gain shifts
over time is what causes the broadening in the pixel gain
distribution.

We correct for these gain shifts on a pixel-by-pixel basis using
energy calibration data from the tagged source as described above.
The spectra in this paper are taken from 8 broad energy bands, and so
are not expected to suffer noticeably from response broadening due to
differing reductions in CZT gain.

\subsection{Cleaning of Bright Sources}

During the course of the BAT survey data processing, skymaps are
generated from each spacecraft pointing and then combined into
mosaicked maps.  In order to reduce the systematic noise present in
the maps from the sidelobes of strong sources, the analysis ``cleans''
bright sources found in the individual snapshot images.  This means
that if a source is detected in a snapshot image with a significance
greater than 6~sigma, we measure its flux and subtract the
contribution of a source with the fitted strength.  After all bright
sources above the cleaning threshold are removed in this way, the
cleaned maps are combined into the mosaicked maps and source detection
performed to locate sources with lower significance.

The data processing pipeline for previous versions of the BAT survey
catalog used a cleaning threshold in the individual snapshot images of
9~sigma.  For the 70~month survey catalog, we have lowered the
threshold to 6 sigma in order to further reduce the systematic noise
in the mosaicked maps due to strong sources.

In addition, we have designated 100 objects as sources that are always
cleaned in every snapshot image, regardless of detected significance.
These are primarily the brightest sources in the survey, usually
galactic and variable.  They sources are always cleaned even if they
are below the 6$\sigma$ threshold for cleaning in a single snapshot
image.  This allows us to remove systematic noise caused by these
strong sources from the mosaicked maps without becoming vulnerable to
unnecessary cleaning of noise peaks as a result of a lower snapshot
cleaning threshold.

\subsection{Pixelization in Mosaicked Sky Maps}

The individual snapshot skymaps are combined into the mosaicked map by
interpolating the pixel values from the snapshot grid onto the mosaic
grid.  In previous versions of the data processing, the mosaic pixel
grid pitch was 5.0~arcminutes (the PSF in the BAT mosaicked images can
be described by a Gaussian with 19.5 arcminute FWHM).  For the
70~month survey we have chosen to use a smaller mosaic grid with a
pixel pitch of 2.8~arcminutes.  This change to smaller mosaic pixels
allows a more accurate determination of source positions and fluxes by
reducing the effects of interpolation error when adding the individual
snapshot images to the mosaicked images.  The choice of a
2.8~arcminute mosaic pixel size results from a compromise between high
precision in the mosaicked maps and reasonable computation times for
the data processing.

\subsection{Crab Weighting}
\label{crabweighting}

In previous iterations of the \swiftbat\ survey, the mosaicked
total-band map (14--195~keV) used for detecting new sources was
produced by simply summing together the individual maps from the eight
energy bands of the survey.  This is equivalent to performing a
weighted sum across the 8 energies with each band having a weight of
unity.

In order to improve the sensitivity of the source detection stage in
the 70~month survey to objects with AGN-like spectra, we have adopted
new weights derived from the measured Crab count rates in the 8 bands.
This has the effect of enhancing detection for sources with a
Crab-like power law spectrum.  Since most AGN have power-law like
spectra with spectral indices of $\sim2$ (close to the Crab's
spectral index of 2.15; see Equation~\ref{eqn:crab_cts}), we expect this
Crab-weighting to improve our detection sensitivity for AGN.

In Crab units, the count rate and noise rate of an individual pixel in
the combined map can be represented as $F_i/K_i$ and $n_i/K_i$ where
$F_i$ is the measured count rate in energy band $i$, $K_i$ is the
measured Crab rate, and $n_i$ is the measured noise rate.  Combining
these individual band measurements in a weighted mean (and dropping
the subscripts) yields
\begin{equation}
F_{cw} = \sum \frac{\left(F/K\right)}{\left(n/K\right)^2}\frac{1}{\sum\left(K/n\right)^2},
\end{equation}
where F$_{cw}$ is the Crab-weighted total count rate in
cts~s$^{-1}$~detector$^{-1}$. This is equivalent to
\begin{equation}
F_{cw} = \sum F\ \frac{\left(K/n^2\right)}{\sum \left(K/n\right)^2}.
\end{equation}
If we express the weighted sum more simply as
\begin{equation}
F_{cw} = \sum_{i=1}^8 W_iF_i,
\end{equation}
then the weights $W_i$ can be constructed with the following formula:
\begin{equation}
W_i = \frac{K_i}{n_i^2 \sum (K_i/n_i)^2}.
\end{equation}

\begin{figure}
  \begin{center}
    \ifpdf
      \resizebox{\columnwidth}{!}{\includegraphics{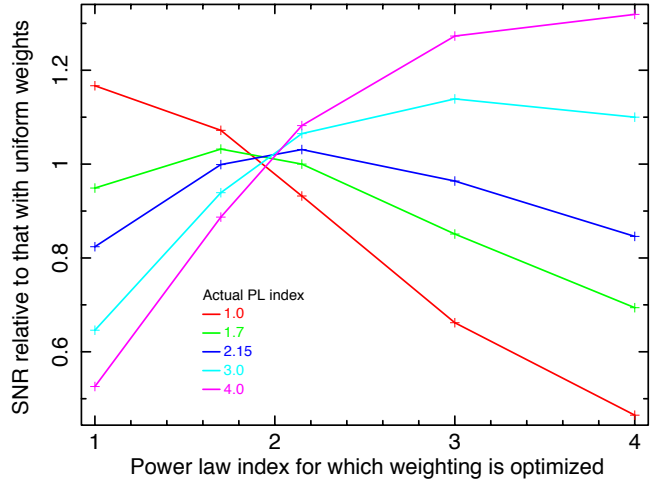}}
    \else
      \resizebox{\columnwidth}{!}{\includegraphics{weighting_v_pl_index.eps}}
    \fi
  \end{center}
\caption{Source detection efficiency as a function of spectral
  weighting.  This figure shows the change in S/N ratio of a source in
  the survey when the input 8-band maps are combined into a total band
  map using weights optimized for a source having a spectrum with a
  particular spectral index.  For example, the blue curve shows that
  when the weighting is optimized for a source with a spectral index
  of 1.0, the measured S/N of a source with an actual spectral index
  of 2.15 is only 80\% as high as when the weighting is uniform
  (weights=1, or simply summing together the 8
  bands).\label{pixellation}}
\end{figure}

Figure~\ref{pixellation} shows the sensitivity of the source detection
process to the spectral slope used to compute the weights for
combining the individual energy bands.
The values for the measured count rate of the Crab in the eight energy
bands, the noise, and the derived values of the weights can be found
in Table~\ref{energy_bands}.

\section{The \swiftbat\ 70~Month Catalog}
\label{catalog}

Table~\ref{table_sources} presents the catalog of sources detected by
\swiftbat\ using the first 70 months of data and includes sources at
all galactic latitudes.  This 70~month catalog and associated data in
electronic form can be found online at the \swiftbat\ 70-month survey
website.

Table~\ref{table_sources} lists all the sources detected above the
4.8$\sigma$ level in a blind search of the 70~month \swiftbat\ survey
maps.  The first column is the source number in the 70~month catalog.
The second column of the table is the BAT name, constructed from the
BAT source position given in columns three and four.  In cases where
the source has been previously published with a BAT name corresponding
to a slightly different location (e.g., a source position from a
previous BAT catalog with less exposure), we have used the first
published name but have given the 70~month BAT coordinates in columns
three and four.  If there is more than one X-ray counterpart to a
single BAT source, we have repeated the BAT name with an ``A'' or
``B'' suffix and used the same BAT coordinates for each of the
counterparts.  The fifth column is the significance of the blind BAT
source detection.  The significance was calculated by taking the flux
at the highest pixel discovered in the blind search in the total-band
mosaic and dividing by the local noise as discussed in
Section~\ref{flux}.  Instances where more than one possible
counterpart to a single BAT source is found within the match radius of
22 arcminutes are indicated with ditto marks in columns 2--5.  We use
the 4.8$\sigma$ significance level as the threshold for inclusion in
the survey as discussed in \cite{tueller22}, since we expect one false
BAT detection on the entire sky at this significance level.

The sixth column gives the name of the counterpart to the BAT hard
X-ray source.  These are often well known X-ray sources, optical
galaxies, or 2MASS sources, and are associated with a source detected
in the medium-energy X-ray band (3--10~keV) in \swiftxrt, \chandra, or
\xmm/ images.  If no counterpart to the BAT source has been
identified, we give the BAT name from column~2 as the counterpart
name.  Counterpart determination is discussed in more detail in
\S\ref{counterparts}.  The seventh column gives an alternate name for
the counterpart; we list a well known name or a name from a hard X-ray
instrument or high energy detection.  The best available coordinates
of the counterparts (J2000) are given in the table in columns~8 and 9.

The 10th and 11th columns give the 14--195~keV flux of the BAT source
(in units of 10$^{-12}$~ergs~sec$^{-1}$~cm$^{-2}$) and the 90\%
confidence interval.  The BAT flux for each counterpart is extracted
from the hard X-ray map at the location of the counterpart given in
columns 8 and 9.  The flux determination method uses a power-law
spectral fit to the flux measurements in each of the 8 energy bands
and is described in \S\ref{flux}.

The 12th column indicates whether there is contamination of the flux
measurement from nearby sources.  This can be the result of source
confusion (two BAT sources close enough together to be unresolved, or
two viable counterparts within the same BAT mosaic pixel), or from the
presence of a strong nearby source.  The number given is the
contamination fraction, or the fraction of the measured flux at the
counterpart position contributed by other sources.  Contamination
fractions are given for all sources with a contamination level greater
than 2\%. The treatment of confused sources and of contaminated flux
measurements is described in more detail in Section~\ref{confused}.

When a source has an entry in column 12 and is considered confused,
the counterpart flux listed in column 10 is an estimate from a
simultaneous fit of all the counterparts in the region to the BAT map.
In these cases, the error on the flux is not well defined and column
11 is left blank. (See \S\ref{flux}).

The 13th and 14th columns list the source spectral index and error in
the BAT band as described in Section~\ref{flux}. The 15th column lists
the reduced $\chi^2$ value from a spectral fit to a power law model
(see \S\ref{flux}).

The 16th and 17th columns give the redshift and BAT luminosity of the
counterpart if it is associated with a galaxy or AGN.  The source
luminosity (with units log[ergs~s$^{-1}$] in the 14--195~keV band) is
computed using the redshift and flux listed in the table and a
cosmology where $H_0 = 70$~km~s$^{-1}$~Mpc$^{-1}$, $\Omega_m = 0.30$,
and $\Omega_\Lambda = 0.70$.

The 18th column can contain a flag indicating the strength
of the association between the BAT source and the listed counterpart
(see \S\ref{ass_strength}).

The 19th column gives an integer source class that we have found
useful for selecting particular classes of objects (e.g., AGN, HMXBs,
etc.).  The source classes are listed in Table~\ref{type-decomp}.

The 20th column lists a source type in the form of a short description
of the counterpart.

\subsection{Counterparts}
\label{counterparts}

As mentioned in Section~\ref{procedure} and described in more detail
in \cite{tueller22}, counterparts to BAT sources were identified by
examining archival X-ray observations from \swiftxrt, \chandra, \xmm,
\suzaku, and \asca.  All point sources near the blind BAT source position
(within $\sim 15$~arcmin
) were checked to see whether an extrapolation of the fit X-ray
spectrum into the BAT band would be consistent with the measured BAT
flux.  Any such X-ray sources with an extrapolated flux above the BAT
detection threshold were checked against SIMBAD and NED to determine a
source name and type and saved to a file containing all discovered
counterparts.

Where there were no archival X-ray observations covering the BAT
source, we submitted the coordinates to the \swift-XRT for a 10~ks
X-ray followup observation.  In cases where we did not yet have X-ray
observations of the BAT source from the X-ray archives or from
\swiftxrt, a best guess was made as to the counterpart by checking for
likely sources (eg, strong, nearby Seyfert galaxies) in NED and
SIMBAD.  These cases are indicated in Table~\ref{table_sources} using
the association strength flag in column~18 as described in
Section~\ref{ass_strength}.




There are a few complications to this procedure.  The blind source
detection technique used on the BAT mosaics (see \cite{tueller22})
does not do a good job of automatically discovering weaker BAT sources
close to very strong sources.  To guard against this case we checked the
BAT mosaics by eye to ensure that all sources detected by BAT are
listed in Table~\ref{table_sources}.

\begin{figure}
  \begin{center}
    \ifpdf
      \resizebox{\columnwidth}{!}{\includegraphics{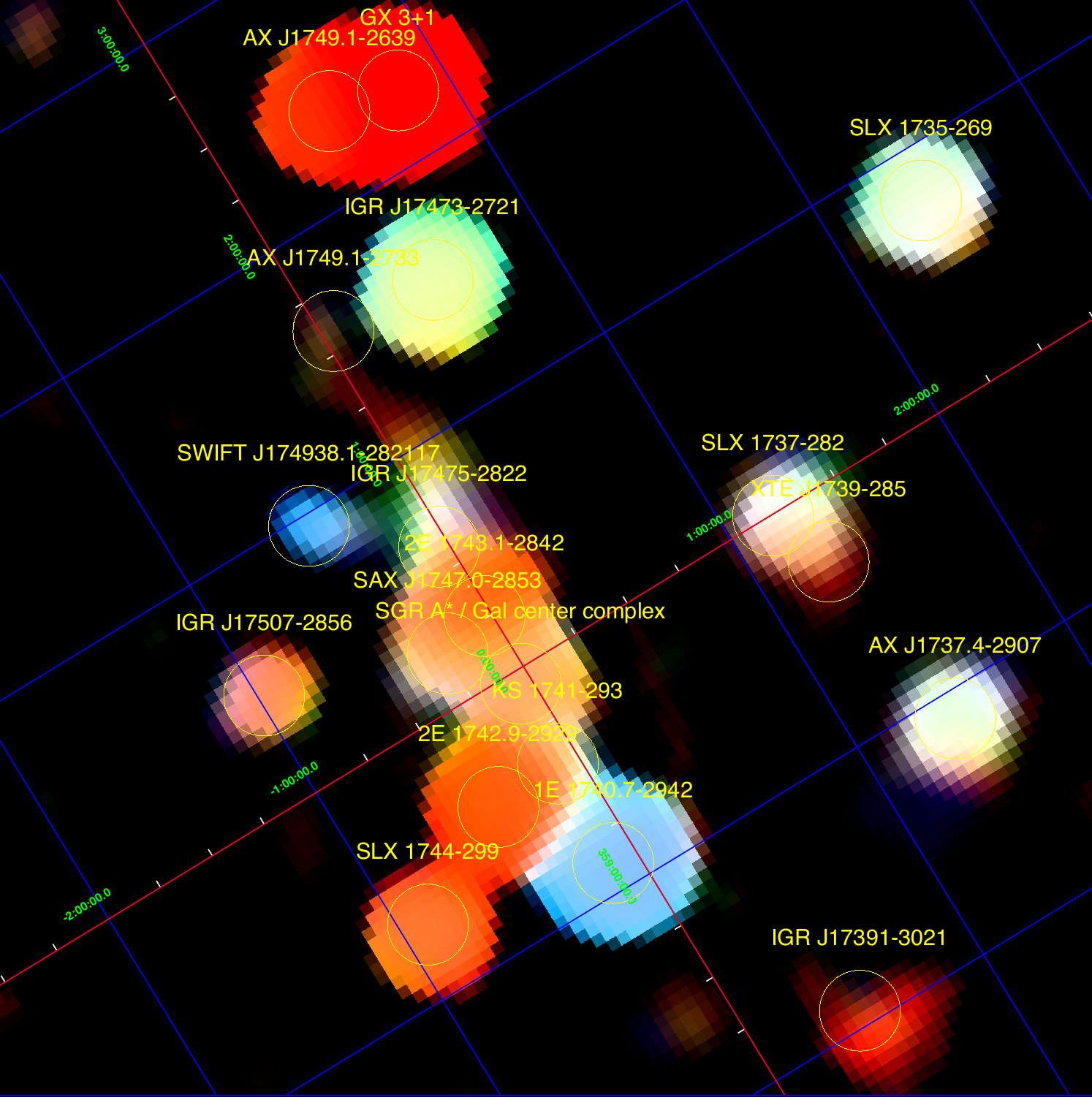}}
    \else
      \resizebox{\columnwidth}{!}{\includegraphics{BAT_58m_galactic_center.ps}}
    \fi
  \end{center}
\caption{\label{gal_center}The galactic center as seen by BAT.  The
  colors of the sources were derived from the 8-band mosaics: the
  lowest two bands (14--24~keV) are combined to give a red value for
  each pixel, 24--50~keV for green, and 50--150~keV for blue. The
  colors are then adjusted to the Crab spectrum, so that white
  sources have a Crab-like spectrum, red sources are softer, and blue
  sources are harder.  The grid is in galactic coordinates.
  The circular regions are centered on the counterparts listed in
  Table~\ref{table_sources} and have a 10~arcminute radius.  The PSF
  of BAT in the mosaicked maps is 19.5~arcminutes.}
\end{figure}

Source detection is especially difficult in the crowded galactic
center region.  Figure~\ref{gal_center} illustrates this problem with
the BAT map of the galactic center region.  The separation between
sources in this crowded region can be on the order of the
19.5~arcminute (FWHM) PSF of the BAT survey.  Given this difficulty
with source detection and confusion in the galactic center region, we
have chosen by hand a reasonable set of the brightest counterparts
needed to explain the measured BAT emission in this area.

\subsubsection{Counterpart associations}
\label{ass_strength}

One of the main goals of the BAT 70 month survey is to produce a uniform
and well identified catalog of hard X-ray sources in the sky.  In
order to achieve this goal we have paid particular attention to the
assignment of counterparts to the sources detected by BAT.  Instead of
just cross-correlating the BAT catalog with catalogs from other
telescopes, we have whenever possible checked the archives for X-ray
observations of the fields containing BAT sources.  We believe that
this level of effort is necessary in order to produce the most useful
catalog for further studies of hard X-ray sources.

An indication of the strength of the association between the BAT
source and the counterpart listed in the main table is given by the
flag listed in Column~18.  This flag is meant to indicate the result
of a check of the X-ray archives for a counterpart to the BAT source.
Table~\ref{table_assoc_strength} lists the possible values for the
association strength flag, their meanings and the number of sources in
each category.

The strongest association is indicated by a value of 0 (blank in the
printed table) in the association strength column.  These sources have
been observed in the medium energy X-ray band (3--10~keV), and images
of the fields checked for possible counterparts to the BAT hard X-ray
source.  Any sources found are checked to see whether their X-ray
spectrum extrapolated into the BAT band yields a flux value above the
BAT survey threshold.  Practically, this is often indicated by the
presence of the source in the medium energy X-ray band since most
sources found in the total X-ray band (0.5--10 keV) are soft and do
not have BAT counterparts.  Over 80\% of the counterparts to sources
in the BAT 70 month survey have been verified with X-ray observations;
although 100\% of the BAT sources have archival X-ray observations,
20\% of the sources have X-ray data that remains to be analyzed.

A value of 1 indicates that the counterpart association has been held
over from a previous catalog, but may not have been explicitly checked
with X-ray archival data.  These are usually very bright or galactic
sources.

A value of 2 or 3 in the association strength column indicates an
intermediate-level association.  A value of 2 means that an
examination of an X-ray image found a plausible soft (1--10keV)
source, but that this source is weak or nonexistent in the hard band
(4--10~keV).  A value of 3 indicates that the evidence for the
counterpart association comes from another waveband, such as an
optical or radio QSO catalog.

A value of 4 in the association strength column means that this source
has archival X-ray data but that it has not yet been checked to verify
the association listed in the catalog.  The association listed is
based on the presence of a nearby object in the SIMBAD or NED
databases likely to be the counterpart to the BAT source.  Such
sources are usually bright Seyfert galaxies or QSOs, or bright
galactic sources.  Past experience with previous BAT catalogs has
shown that counterpart associations made in this way and later
verified with X-ray observations are correct greater than 95\% of the
time.

A value of 5 or 6 in the association strength column indicates that
the field has been observed in the X-ray band but that no counterpart
was found.  This usually indicates that the BAT source is transient
and not detected in the X-ray observation.  Sources on the galactic
plane are indicated with a 5, and sources greater than 10\degree from
the galactic plane are indicated with a 6.

A value of 7 in the association strength column means that the source
has no archival data for observations in other wavebands and that no
suitable counterpart has been found.

%
%
\begin{center}
\begin{deluxetable}{crrp{4.9cm}}
\tablecaption{Counterpart Association Strengths\label{table_assoc_strength}}
\tablewidth{0pt}
\tablehead{
\colhead{Flag} & \colhead{\# in catalog} & \colhead{\%} & \colhead{Meaning}
}
\startdata
(blank) & 1017 & 84.2 & Confirmed with X-ray imaging \\
1       &   36 &  3.0 & Old association held over from previous catalog\\
2       &    5 &  0.4 & No good hard X-ray source; soft X-ray source\\
3       &    4 &  0.3 & No X-ray source; source from another waveband\\
4       &  133 & 11.0 & Unchecked or unavailable X-ray image; educated guess \\
5       &    2 &  0.2 & No X-ray source; BAT source on Galactic plane\\
6       &   10 &  0.8 & No X-ray source; BAT source off plane\\
7       &    3 &  0.2 & No association 
\enddata

\end{deluxetable}
\end{center}

\subsection{Source Types and Distribution}

%
%
\begin{deluxetable}{rlr}
\tablecaption{Counterpart Types in the \swiftbat\ 70~month Catalog\label{type-decomp}}
\tablewidth{\columnwidth}
\tablehead{
\colhead{Class} & \colhead{Source Type} &
\colhead{\# in catalog}
}
\startdata
0       & Unknown\tablenotemark{a}              & 65 \\
1       & Galactic \tablenotemark{b}            & 23 \\      
2       & Galaxy\tablenotemark{c}               & 111 \\
3       & Galaxy Cluster                        & 19 \\
4       & Seyfert I (Sy1.0--1.5)                & 292 \\
5       & Seyfert II (Sy1.7--2.0)               & 261 \\
6       & Other AGN                             & 23 \\
7       & Blazar / BL Lac                       & 49 \\
8       & QSO\tablenotemark{d}                  & 86 \\
9       & Cataclysmic Variable star (CV)        & 55    \\
10      & Pulsar                                & 20    \\
11      & Supernova Remnant (SNR)               & 6    \\
12      & Star                                  & 14    \\
13      & High Mass X-ray Binary (HMXB)         & 85    \\
14      & Low Mass X-ray Binary (LMXB)          & 84    \\
15      & Other X-ray Binary (XRB)            & 17    \\
        &                      & --- \\
        & Total                & 1210

\enddata
\tablenotetext{a}{Sources listed with the type unknown either do not have
  any known counterpart, or are associated with sources of unknown
  physical type.}
\tablenotetext{b}{Sources classified only as ``Galactic'' are so assigned
  because of observed transient behavior in the X-ray band along with
  insufficient evidence to place them in another class.}
\tablenotetext{c}{Sources in the ``Galaxy''  class are seen as
  extended in optical or near-IR imagery, but do not have firm
  evidence (such as an optical spectrum) from other wavebands
  confirming whether they harbor an AGN.}  
\tablenotetext{d}{AGN with BAT luminosities greater than
  $10^{45}$~ergs~sec$^{-1}$ or listed in NED as radio galxies are
  classified as QSO.}
\end{deluxetable}

\begin{figure*}
\begin{center}
  \ifpdf
    \resizebox{0.99\textwidth}{!}{\includegraphics{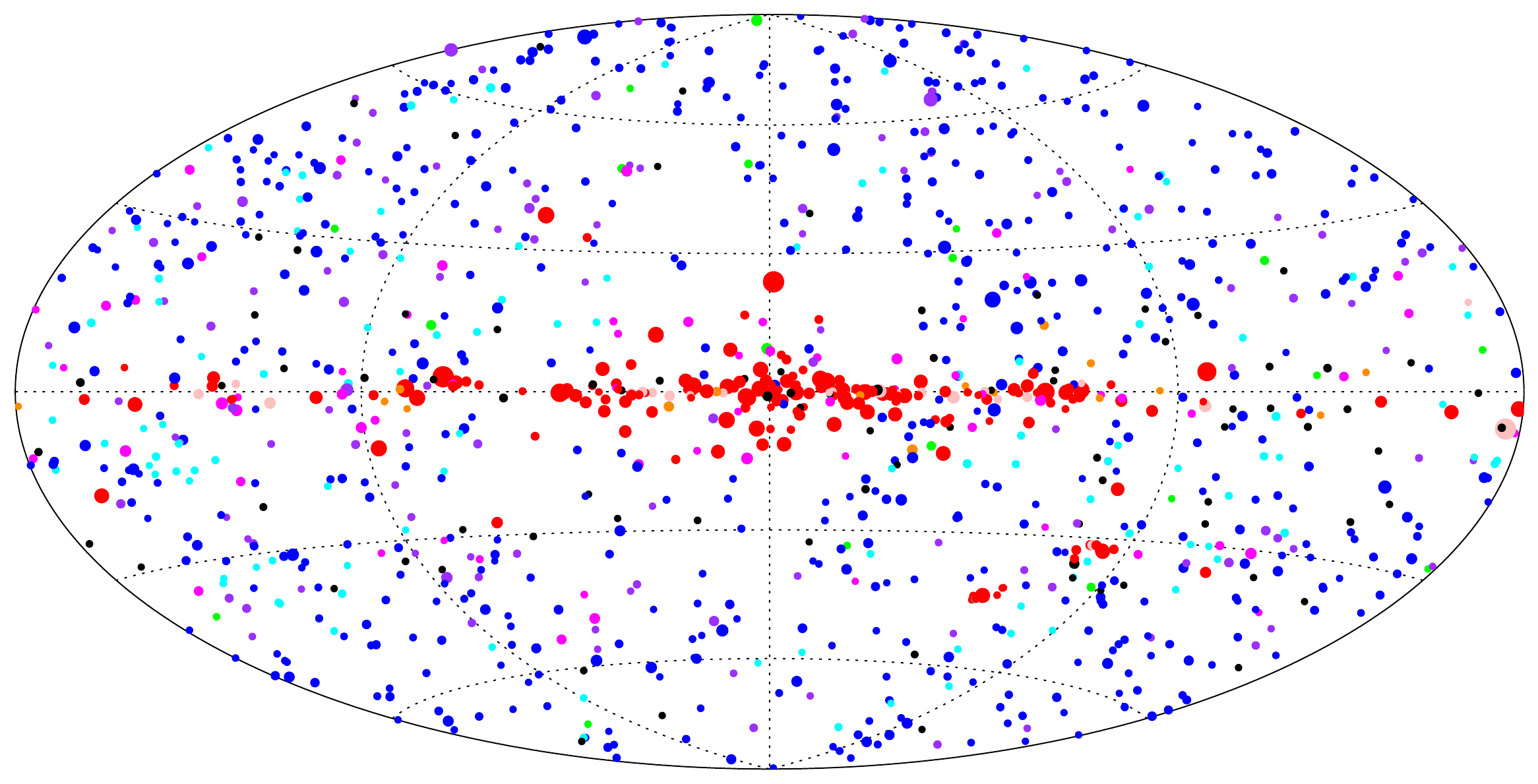}}
    \resizebox{0.99\textwidth}{!}{\includegraphics{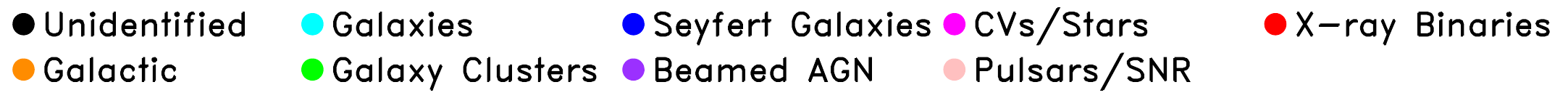}}
  \else
    \resizebox{\textwidth}{!}{\rotatebox{90}{\includegraphics*[.8in,.8in][6in,10in]{70m_sourcetype_allsky.ps}}}
    \resizebox{\textwidth}{!}{\rotatebox{90}{\includegraphics*[2.3in,.8in][2.8in,9.8in]{70m_legend_sourcetype.ps}}}
  \fi
\caption{All-sky map showing classification of the BAT 70~month survey
  sources.  The figure uses a Hammer-Aitoff projection in galactic
  coordinates; the flux of the source is proportional to the size of
  the circle.  The source type is encoded by the color of the
  circle.\label{aitoff-type}}
\end{center}
\end{figure*}

Figure~\ref{aitoff-type} shows the distribution of sources on the sky
color~coded by source type, with the symbol size proportional to the
source flux in the 14--195~keV band.  Table~\ref{type-decomp} gives
the distribution of objects according to their source type.  Sources
classified as ``unknown'' are those where the physical type of the
underlying object (e.g., AGN, CV, XRB, etc) has not yet been
ascertained.  These sources often have a primary name derived from the
BAT position.  Some BAT sources of unknown type are associated with named
sources discovered by other missions in the X-ray or gamma-ray bands,
but are classified as unknown in Table~\ref{table_sources}
because the physical type of the named source is unknown or because
the coordinates of the source are not precise enough to identify an
optical counterpart.  These sources can be distinguished by having a
name in the catalog derived from the observation in the other
waveband.  The few sources classified only as ``Galactic'' generally
lie in the plane and have shown some transient behavior which
indicates a galactic source, but no other information is available
that would allow further classification. Sources labeled ``Galaxy''
are detected as extended sources in optical or near-IR imaging, but do
not yet have spectroscopic evidence of being an AGN.

\subsection{BAT Fluxes and Spectra}
\label{flux}

The fluxes of the counterparts to BAT sources were extracted in the
eight BAT bands from the mosaicked maps using the pixel containing the
position of the identified counterpart. For sources where a
counterpart is not known, we use the fitted BAT position to extract
the flux in the eight bands.  The errors associated with the flux
values were calculated by computing the rms value in the mosaicked
maps in an area around each source with a radius of 100 pixels (4.5
degrees).  An exclusion zone around the source with radius of 15
pixels (40.5 arcminutes) was applied to the central source and any
nearby sources that fell in the background calculation area.

We
normalize the measured fluxes in each band to the Crab as described in
\cite{tueller22} by using the equation
\[
\mathrm{BAT\;source\;flux = \left(\frac{BAT\;source\;count\;rate}
{Crab\;count\;rate}\right) Crab\;flux }.
\]
We take the Crab counts spectrum to be
\begin{equation}
\label{eqn:crab_cts}
F(E) = 10.17\:E^{-2.15} 
\left(\frac{\rm photons}{{\rm cm}^2\,{\rm sec}\:{\rm keV}}\right),
\end{equation}
(see \cite{tueller22}).  The total Crab flux is then
\begin{equation}
\mathrm{Crab}\;\mathrm{flux} = \int_{14\;keV}^{195\;keV} 
E\;F(E)\:dE = 2.386\times 10^{-8}
\left(\frac{\rm ergs}{{\rm cm}^2\,{\rm sec}}\right).
\label{crab_flux_eqn}
\end{equation}

After normalizing the 8 band flux values to the Crab, the BAT spectra
are then packed into standard pulse height analysis fits files with
the appropriate keywords for spectral fitting.

As in \cite{tueller22}, we fit these 8-channel spectra with a power
law model in order to find the flux of each source.  We use a BAT
spectral response matrix generated by the BAT software
\texttt{batdrmgen} that has been Crab normalized to be compatible with
the BAT spectra.  We use the \texttt{XSPEC} fitting software to fit
the 8-channel spectra with the \texttt{pegpwrlw} model (power law with
pegged normalization) over the 14--195~keV BAT survey energy range.
We list the spectral index and flux determined from the fit in
Table~\ref{table_sources}.

The 90\% confidence intervals for the overall flux and the spectral
index were found by using the \texttt{error} function in
\texttt{XSPEC} and are given in Table~\ref{table_sources}.  For the
highest significance BAT sources ($>100$~sigma), this procedure does
not produce a good fit (reduced $\chi^2 \gg 1$). However, this is to
be expected from the very high significance of each data point, the
coarse energy binning, and because a simple power law is not a good
model for the spectra of many galactic objects.  We list the reduced
$\chi^2$ for each source in Table~\ref{table_sources} as an indicator
of which sources are not well fit with a power law model, but leave a
more detailed spectral analysis to a later work.

We present the eight band BAT spectra themselves in
Table~\ref{table_spectra}.  The printed version of this table lists
only a few objects for space reasons.  However, the full table listing
all the sources can be found in an electronic version on the ApJ
website.  We also provide pha fits files for all the BAT spectra and
an ASCII table at the \swiftbat\ survey
website.
\tabletypesize{\scriptsize}
\tabletypesize{\tiny}
\begin{deluxetable*}{llrrrrrrrrrrrrrrrrr}
\setlength{\tabcolsep}{0.06in}
\tablecaption{Spectra of Sources in the 70 month \swiftbat\ Survey \label{table_spectra}}
\tablewidth{0pt}
\tablehead{
\colhead{BAT Name}& \colhead{Counterpart Name}& \colhead{C\tablenotemark{a}}& \colhead{14--20}& \colhead{Err\tablenotemark{b}}& \colhead{20--24} & \colhead{Err\tablenotemark{b}}& \colhead{24--35} & \colhead{Err\tablenotemark{b}}& \colhead{35--50} &\colhead{Err\tablenotemark{b}}& \colhead{50--75} &\colhead{Err\tablenotemark{b}}& \colhead{75--100} &\colhead{Err\tablenotemark{b}}& \colhead{100--150} & \colhead{Err\tablenotemark{b}}& \colhead{150--195} &\colhead{Err\tablenotemark{b}}\\\colhead{}& \colhead{}& \colhead{}& \colhead{keV\tablenotemark{b}}& \colhead{}& \colhead{keV\tablenotemark{b}}& \colhead{}& \colhead{keV\tablenotemark{b}}& \colhead{}& \colhead{keV\tablenotemark{b}}& \colhead{}& \colhead{keV\tablenotemark{b}}& \colhead{}& \colhead{keV\tablenotemark{b}}& \colhead{}& \colhead{keV\tablenotemark{b}}& \colhead{}& \colhead{keV\tablenotemark{b}}& \colhead{} 
}
\startdata
SWIFT J0003.3$+$2737  & 2MASX J00032742+2739173   &       &   2.72 &  1.74 &    1.91 &  0.84 &    4.06 &  1.24 &    3.34 & 0.87 &    1.93 &  0.83 &    1.46 &  0.74 &   -0.33 &  1.31 &   -3.24 &  2.72 \\
SWIFT J0005.0$+$7021  & 2MASX J00040192+7019185   &       &   7.56 &  1.57 &    2.90 &  0.75 &    4.36 &  1.12 &    3.03 & 0.76 &    2.24 &  0.77 &    0.73 &  0.70 &   -0.72 &  1.21 &   -3.38 &  2.48 \\
SWIFT J0006.2$+$2012  & Mrk 335                   &       &  12.80 &  1.72 &    4.98 &  0.85 &    6.59 &  1.27 &    3.88 & 0.85 &    2.54 &  0.85 &    0.97 &  0.77 &    1.69 &  1.34 &    3.89 &  2.89 \\
SWIFT J0009.4$-$0037  & 2MASX J00091156-0036551   &       &   5.36 &  1.88 &    2.58 &  0.92 &    2.26 &  1.40 &    1.90 & 0.92 &    1.76 &  0.87 &    0.26 &  0.79 &    1.64 &  1.48 &   -2.54 &  3.06 \\
SWIFT J0010.5$+$1057  & Mrk 1501                  &       &  14.00 &  1.83 &    5.15 &  0.88 &   10.30 &  1.29 &    5.25 & 0.85 &    6.41 &  0.87 &    1.95 &  0.77 &    2.52 &  1.43 &    1.48 &  2.92 \\
SWIFT J0017.1$+$8134  & [HB89] 0014+813           &       &   8.64 &  1.62 &    3.05 &  0.75 &    3.61 &  1.16 &    2.51 & 0.77 &    0.06 &  0.72 &    1.39 &  0.65 &    2.23 &  1.19 &    4.01 &  2.46 \\
SWIFT J0021.2$-$1909  & 2MASX J00210753-1910056   &       &   7.99 &  1.81 &    3.58 &  0.84 &    7.50 &  1.26 &    3.52 & 0.91 &    2.73 &  0.86 &    0.70 &  0.80 &    0.86 &  1.41 &   -2.64 &  2.95 \\
SWIFT J0023.2$+$6142  & IGR J00234+6141           &       &   8.74 &  1.67 &    3.59 &  0.75 &    5.29 &  1.14 &    1.50 & 0.80 &   -1.02 &  0.74 &    0.40 &  0.66 &    1.28 &  1.22 &    0.88 &  2.52 \\
SWIFT J0025.2$+$6410  & Tycho SNR                 &       &  17.70 &  1.67 &    5.77 &  0.76 &    3.96 &  1.14 &    1.49 & 0.78 &    1.03 &  0.75 &    1.85 &  0.66 &    1.82 &  1.21 &    5.44 &  2.54 \\
SWIFT J0025.8$+$6818  & 2MASX J00253292+6821442   &       &   7.71 &  1.58 &    2.14 &  0.77 &    3.64 &  1.14 &    3.45 & 0.77 &    3.34 &  0.76 &    1.39 &  0.70 &    2.43 &  1.21 &    2.06 &  2.55 
\enddata
\tablecomments{Table~\ref{table_spectra} is published in its entirety in the electronic edition of this atrticle.  A portion is shown here for guidance regarding its form and content.}
\tablenotetext{a}{Contamination fraction (see \S\ref{confused})}
\tablenotetext{b}{[$10^{-5}$cts\,s$^{-1}$]}
\end{deluxetable*}
\renewcommand{\arraystretch}{1.0}

\begin{figure*}
  \begin{center}
    \ifpdf
      \resizebox{\textwidth}{!}{\includegraphics{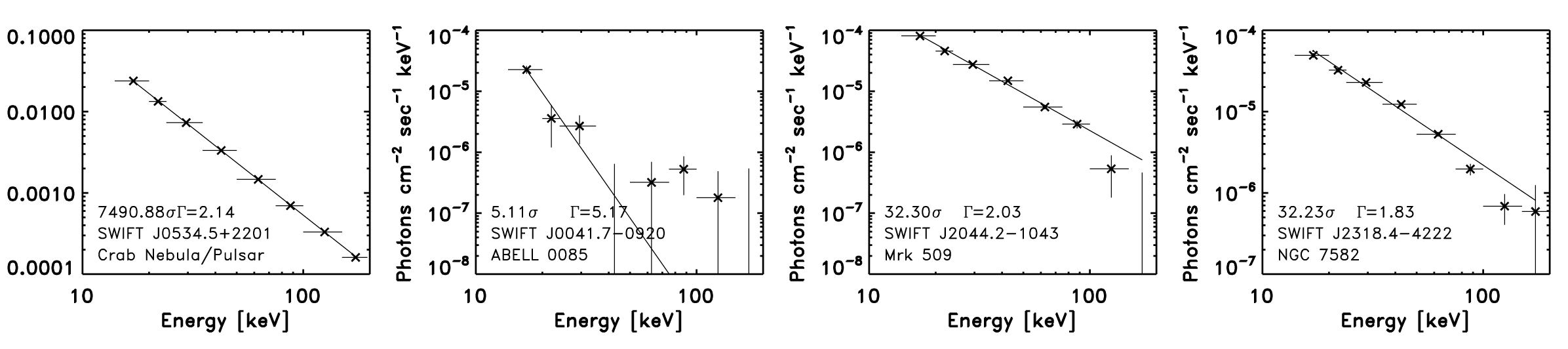}}
    \else
      \resizebox{\textwidth}{!}{\includegraphics{58m_spectra_poster.eps}}
    \fi
  \end{center}
\caption{\label{spectra}BAT spectra of four sources in the 70~month
  catalog.  The line in the plots is a simple power law fit to the
  data, with the power law index given as $\Gamma$ in the legend.}
\end{figure*}

Figure~\ref{spectra} shows the spectra of four representative sources
from the BAT 70~month survey. 

\subsection{Confused Sources}
\label{confused}

Sources are labeled as confused in our table (i.e., they have a value
in the contamination fraction column of Table~\ref{table_sources})
when the highest pixel associated with the BAT source in the mosaicked
maps (the ``central pixel'' value) has a significant contribution from
adjacent sources. This includes the cases when two possible X-ray
counterparts lie within a single BAT pixel and when two BAT sources
are close enough that each contributes flux at the location of the
adjacent source.

Using the positions of the X-ray counterparts as an input catalog, we
determined the contamination fraction of each source by using the
mosaicked total band map to simultaneously fit for the intensity of a
PSF (a 19.5 arcminute FWHM Gaussian) centered at the location of each
source.  Using these fit values, we decompose the measured flux at
each source location into a contamination rate contributed from nearby
sources and a fit rate for the central source.  We then compute a
contamination fraction,
\[\mathrm{Contamination~Fraction} = \frac{\mathrm{Contamination~rate}}{\mathrm{Fit~rate}}. \]
If the source has a contamination fraction greater than 2\%, we list
the value in Table~\ref{table_sources} and consider the source
confused.

The fluxes for sources marked as confused were calculated in a
slightly different way than for unconfused sources.  Instead of using
the measured count rate extracted from the map at the counterpart
position, we use the decomposed fit rate extracted in each energy band
to form the source spectrum from which the flux is calculated.  For
these sources we do not quote an error on the flux estimate because
the errors produced with this fitting technique are not well behaved.
Any source with a contamination fraction greater than 2\% can be
considered as detected by BAT, but the quoted flux should be
considered an upper limit.

\subsection{Lightcurves}

Another important goal of this paper survey is to make available
lightcurves that span the 70~month period of the survey for the
sources detected by BAT.  In order to achieve this we have constructed
all-sky total-band mosaic images for each month of data in the survey.
We extract from the monthly mosaics fluxes for all the sources
detected in the full survey and use these to construct monthly sampled
lightcurves that cover the duration of the survey.

\begin{figure*}
  \begin{center}
    \ifpdf
      \resizebox{\textwidth}{!}{\includegraphics{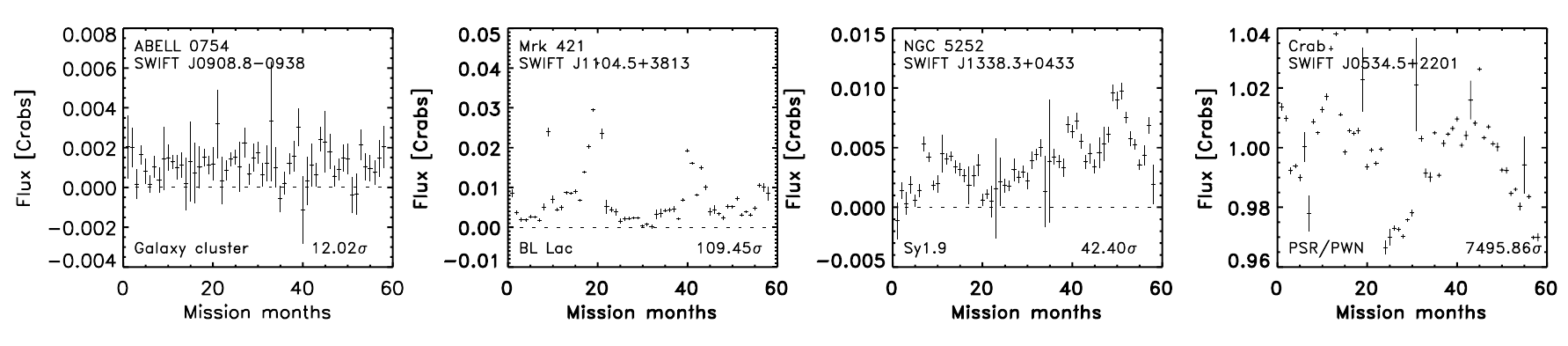}}
    \else
      \resizebox{\textwidth}{!}{\includegraphics*[0.6in,8.7in][7.7in,10.1in]{58m_lightcurves_poster_2.ps}}
    \fi
  \end{center}
\caption{\label{lightcurves}Lightcurves of four sources from the
  70~month catalog.}
\end{figure*}

Figure~\ref{lightcurves} shows four representative lightcurves from
the BAT 70~month survey.
The fourth panel of the figure shows the lightcurve of the Crab.  The
error bars on the data points are constructed from the local noise in
the monthly mosaic image and are representative of the statistical
error associated with each data point.  The five points with very
large error bars result from times when the Sun is near the Crab's
position in the sky and \swift\ is constrained against pointing in
that direction, resulting in smaller exposure times and larger error
bars.

\cite{wilson-hodge} has shown that the flux of the Crab in the BAT
band is not constant, but shows variations of ~10\% over timescales of
order one year.  This behavior matches what we see in the BAT
lightcurve of the Crab.  As in \cite{wilson-hodge}, we estimate the
systematic errors in the BAT lightcurves to be $\sim0.75$\% of the
source flux by assuming that the long-term (months-to-years)
variations in the Crab light curve are due to real variations in the
Crab, and that the shorter term variations around that trend are
representative of the systematic error in the measurements.

The lightcurves for each source detected in the survey can be found at
the
\swiftbat\ website.

\section{Survey Characterization}

\subsection{Source Positional Uncertainty}
\label{position_error}

\begin{figure}
\begin{center}
  \ifpdf
    \resizebox{\columnwidth}{!}{\includegraphics{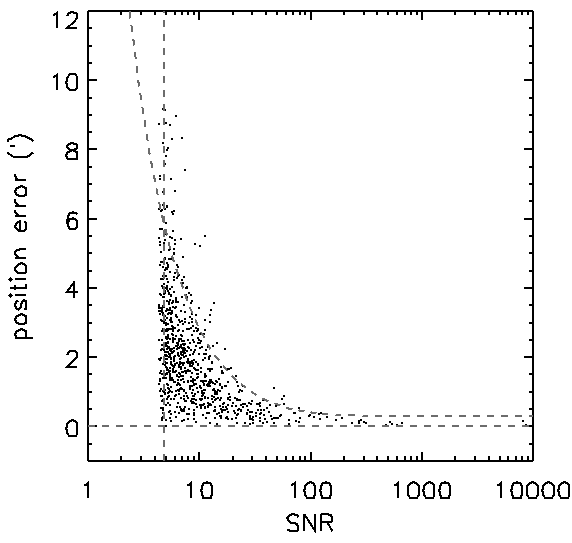}}
  \else
    \resizebox{\columnwidth}{!}{\includegraphics{poserr.eps}}
  \fi
\end{center}
\caption{The BAT position error as a function of the BAT detection
  significance. The angular separation between the counterpart
  position and the fitted BAT position is used to determine a measured
  position error for each source.  Sources closer than $5^\circ$ from
  the galactic plane and galaxy clusters were not included.  The
  measured position error is plotted as a function of BAT detection
  significance.  The dashed line in the plot shows the 91\% error
  radius as a function of BAT source detection significance
  (see~\S~\ref{position_error}).  
  \label{ang_sep}}
\end{figure}

In order to judge the accuracy of the BAT positions, we plot in
Figure~\ref{ang_sep} the angular separation between the BAT position
and the counterpart position against the significance of the BAT
source detection.
In Figure~\ref{ang_sep} we also plot a line showing our estimate of
the BAT position uncertainty for a given source significance.  This estimate
for the error radius (in arcminutes) can be represented with the
function
\begin{equation}
\mathrm{BAT~error~radius (arcmin)} = 
\sqrt{\left(\frac{28}{\left(S/N\right)}\right)^2 +
  \left(0.3\right)^2},
\end{equation}
where $S/N$ is the BAT detection significance.  This empirical
function includes a systematic error of 0.3~arcmin deduced from the
position errors of very significant sources. This systematic error is
consistent with differential aberration across the very large BAT FOV
and with small offsets caused by the slightly energy dependent focal
length of the BAT.  91\% of the BAT sources not on the galactic plane
($l > 5^\circ$) have counterparts that are within this position error
radius.

\subsection{Measured and Theoretical Sensitivity and Systematic Errors}
\label{syserr}

In this section we compare the expected errors with the actual
measured noise in the final mosaic maps.

\subsubsection{Exposure and Noise and Systematic Errors}
\label{exposure}

\begin{figure}
  \begin{center}
    \ifpdf
      \resizebox{\columnwidth}{!}{\includegraphics{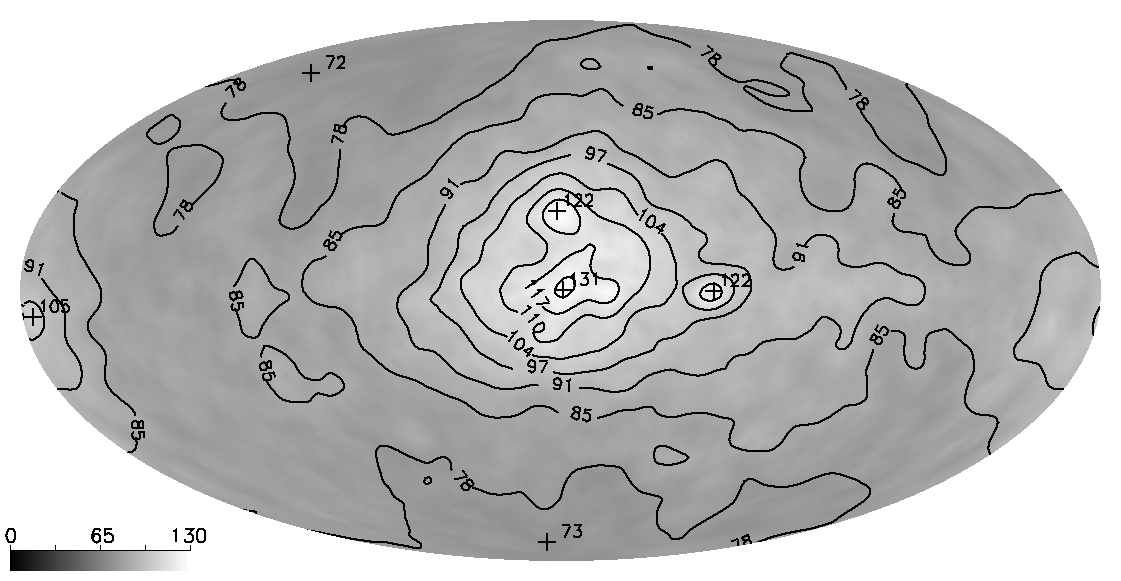}}
    \else
      \resizebox{\columnwidth}{!}{\includegraphics{bkgvar_map_expo_corr.eps}}
    \fi
  \end{center}
\caption{Systematic noise map for the BAT 70~month survey in galactic
  coordinates.  The contours are in arbitrary units.  The measured
  noise map was exposure corrected in order to highlight the
  systematic error contributions.  The highest noise levels (and
  lowest sensitivity) are in the galactic center area which contains
  many bright sources.}
\label{noise_map}
\end{figure}

The observed noise in the 70~month survey can be considered to contain
a primary component related to the statistical uncertainty (dominated
by the exposure time), and a systematic component caused by things
like the incomplete cleaning of bright sources from the maps.
Figure~\ref{noise_map} shows the exposure-corrected noise map for the
70~month survey.  
With the statistical component of the noise removed,
the systematic noise component can be seen concentrated in the center
of the galactic plane near the location of many of the brightest
sources in the sky.  This suggests that the dominant contribution to
the systematic noise is from incomplete cleaning of bright sources.


\subsubsection{Predicted Noise}

From the perspective of pure Poisson counting statistics, the
uncertainties are governed primarily by the properties of the coded
mask and the background (see \cite{skinner08} for details).  The
expected $5\sigma$ noise level can be expressed as (adapting from
\cite{skinner08} Eqn.~23 and 25):
\begin{equation}
5\sigma_{\rm Poisson} = 5 \sqrt{\frac{2b}{\alpha\:N_{\rm det}\:T}},
\label{eq:poinoise}
\end{equation}
where $b$ is the per-detector rate, including background and point
sources in the field of view; $N_{\rm det}$ is the number of active
detectors ($N_{\rm det} \le 32768$); $T$ is the effective on-axis
exposure time; and $\alpha$ is a coefficient dependent on the mask
pattern and detector pixel size ($\alpha = 0.27$ for BAT).\footnote{In
  \cite{tueller22} a typographical error incorrectly listed $\alpha = 0.733$ for BAT.}
The partial coding, $p$, enters the expression through the ``effective
on-axis exposure'' time, $T = p T_o$, where $T_o$ is the actual
exposure time.  Using the measured and exposure weighted values for
the background in each band and the Crab weights given in
Table~\ref{energy_bands},
$b=0.246$~cts~s$^{-1}$~detector$^{-1}$; $N_{\rm det} = 22,520$ (the
exposure-weighted mean number of enabled detectors); and the measured
count rate of the Crab from the mosaics\footnote{In \cite{tueller22},
  we used a value for the Crab count rate of $4.59\times
  10^{-2}$~cts~s$^{-1}$~detector$^{-1}$.  However, this rate was
  incorrectly based on data taken from the individual snapshot images
  instead of from the mosaics.} of $3.86\times
10^{-2}$~cts~s$^{-1}$~detector$^{-1}$ , we find the estimated Poisson
$5\sigma$ noise flux level to be
\begin{equation}
\label{noise_eqn}
f_{5\sigma} = 1.18\:{\rm mCrab} \left(\frac{T}{1\:{\rm Ms}}\right)^{-1/2}.  
\end{equation}


The values for $b$ and $N_\mathrm{det}$ have been measured in the
70~month survey as opposed to only being estimated in the 22~month
survey paper, so we believe we have improved our computation of the
expected noise.

Using the median exposure time of 9.45~Ms shown in
Figure~\ref{exposure_fig}, we use Equation~\ref{noise_eqn} to obtain
the expected median value for the $5\sigma$ sensitivity of the
70~month survey of 0.38~mCrab, or 9.2~$\times
10^{-12}$~ergs~sec$^{-1}$~cm$^{-2}$ using
Equation~\ref{crab_flux_eqn}.

\subsubsection{Measured Noise}

\begin{figure*}
\begin{center}
  \ifpdf
    \resizebox{0.4\textwidth}{!}{\includegraphics{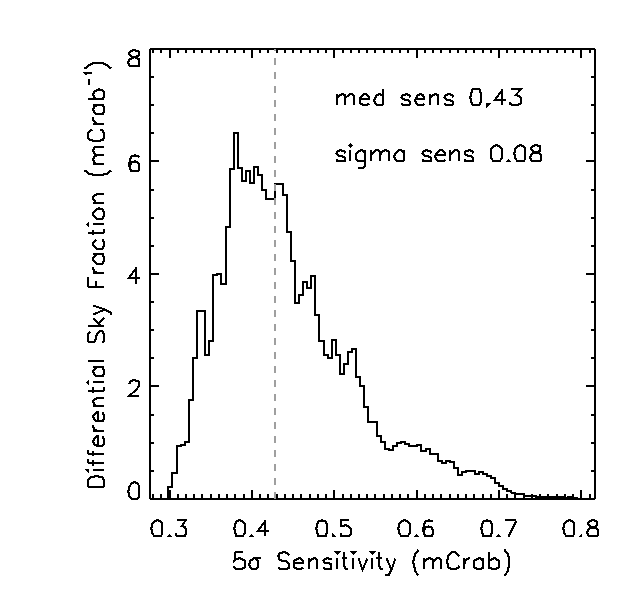}}
    \hspace{.6in}
    \resizebox{0.41\textwidth}{!}{\includegraphics{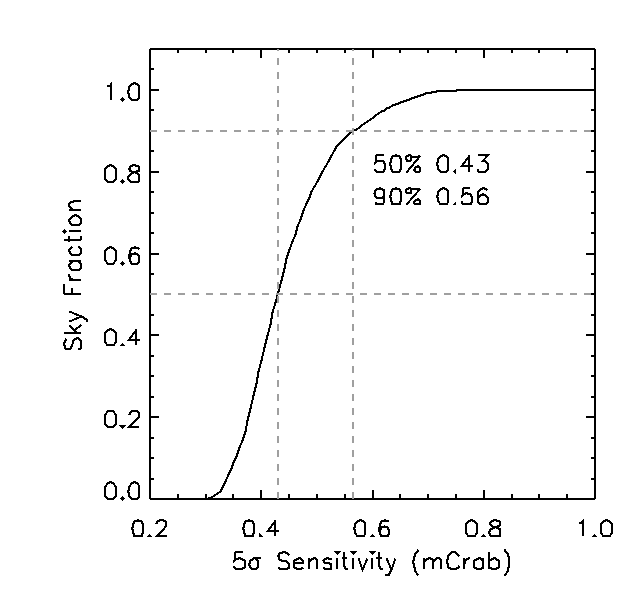}}
  \else
    \resizebox{0.4\textwidth}{!}{\includegraphics{bkgvar_hist.eps}}
    \hspace{.6in}
    \resizebox{0.41\textwidth}{!}{\includegraphics{bkgvar_int.eps}}
  \fi
\end{center}
\caption{Sky coverage versus sensitivity
  achieved in the survey.  The 0.43~mCrab sensitivity limit (for 50\% sky
  coverage) corresponds to a flux of $1.03 \times
  10^{-11}$\,ergs\,cm$^{-2}$\,s$^{-1}$ in the 14--195~keV band.
\label{sens_sky}}
\end{figure*}

The sensitivity of the \swiftbat\ survey is determined by the noise in
the all-sky mosaic maps.  In order to measure the noise and determine
the sensitivity we use the method described in \S\ref{flux} of
calculating the local rms level from the maps in the area around each
source. The sensitivity is then calculated in Crab units using the
measured count rate of the Crab given in Table~\ref{energy_bands}.
Figure~\ref{sens_sky} shows the distribution of sensitivities measured
in the pixels from the all-sky mosaicked map.
The median 5$\sigma$ sensitivity achieved in the 70~month survey is
0.43~mCrab, or $1.0 \times 10^{-11}$\,ergs\,cm$^{-2}$\,s$^{-1}$ in the
14--195~keV band.

We can now compare the measured sensitivity achieved in the survey to
the predicted level computed with Equation~\ref{noise_eqn}.  Taking
the ratio of the 0.38~mCrab median predicted sensitivity to the
0.43~mCrab median measured sensitivity ($0.43 / 0.38 = 1.13$), we find
that the BAT 70~month survey achieves a sensitivity level within 13\%
of the theoretically ideal performance.  In comparison, the BAT
22~month survey achieved a sensitivity within 40\% of expectations.
This advance in achieved sensitivity between the 22 and 70~month
surveys can be attributed to the improvements in the survey data
reduction and processing described in Section~\ref{procedure}.

\begin{figure}
\begin{center}
  \ifpdf
    \resizebox{\columnwidth}{!}{\includegraphics{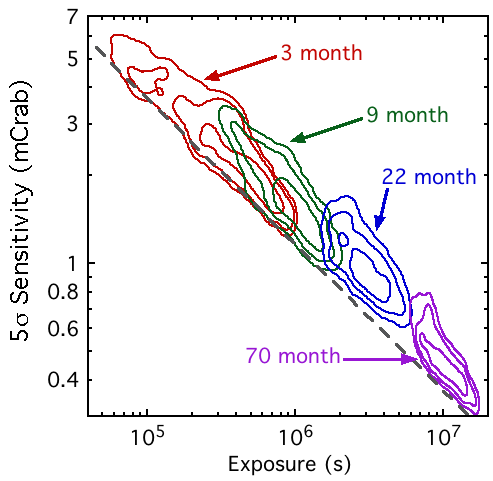}}
  \else
    \resizebox{\columnwidth}{!}{\includegraphics{sensitivity_v_expo.eps}}
  \fi
\caption{Measured $5\sigma$ BAT sensitivity limit for pixels in the
  all-sky map, as a function of effective exposure time, $T$, for the
  3~month (red; \cite{markwardt05}), 9~month (green; \cite{tueller9}),
  22~month (blue; \cite{tueller22}), and 70~month (purple) survey
  analyses.  The contours are linearly spaced and indicate the number
  of pixels with a given sensitivity and effective exposure.  The
  black dashed line represents a lower limit to the expected Poisson
  noise level (see \S\ref{syserr}).  The median achieved sensitivity is
  within 13\% of the predicted value.
  \label{sens_expo} }
\end{center}
\end{figure}

Figure~\ref{sens_expo} shows the relationship between the measured
sensitivity and the predicted sensitivity for all four installments of
the \swiftbat\ survey.  
The black dashed line in the figure gives the theoretical survey
sensitivity as predicted by Equation~\ref{noise_eqn}.  The contours
show the measured sensitivities for the pixels in the all-sky map.
The red contours are from the 3~month survey, the green contours from
the 9~month survey, the blue contours from the 22~month survey, and
the purple contours from this work.  The 70~month contours are much
closer to the dashed predictions than the 22~month contours, again
showing the improvements made in the 70~month data reduction and
processing.  The small tails in the contours at lower exposure and
sensitivity are from areas near the galactic center suffering from
higher systematic noise.

The difference between the theoretical sensitivity and the
measurements increases slightly at longer times (for the 3, 9, and
22~month samples) because the systematic errors are not decreasing
with time the same way that the statistical errors are.  But, the 70
month measured data shows sensitivity closer to the theoretical
expectation. This is because the newly implemented data processing
pipeline for the 70~month survey improves the sensitivity and
reduces systematic errors resulting from problems like uncorrected
pixel gain shifts.

\section{Conclusions}

The \swiftbat\ 70~month catalog is the fourth published catalog
compiled from sources detected in the \swiftbat\ all-sky hard X-ray
survey.  This most recent version of the official \swiftbat\ survey
catalog contains 1171 sources detected from across the entire sky
associated with 1210 counterparts and is the deepest uniform hard
X-ray survey ever conducted.

With detections of over 600 AGN in the hard X-ray band, the
\swiftbat\ 70~month survey catalog contains a valuable reference set
of active galaxies in the local universe.  In addition to the survey
catalog, the database of hard X-ray spectra and lightcurves from
throughout the \swift\ mission will be an important source of
information for future studies of galactic hard X-ray sources and AGN.

\acknowledgements

We would like to acknowledge the help of Mike Koss in obtaining
optical spectra and redshifts for many sources in the table.  This
work has made heavy use of the NED, SIMBAD, and the HEASRAC online
databases as well as the private Leicester database of automatic
analyses of XRT data for the followup observations of BAT survey
sources.  And of course, this work could not have been completed
without the diligent efforts of all members of the \swift\ team.




\setlength{\tabcolsep}{1.2pt}

\clearpage
\LongTables

\renewcommand{\arraystretch}{1}
\tabletypesize{\tiny}
\begin{deluxetable}{p{0.4cm}lrrrllrrrcrrrrccccl}
\setlength{\tabcolsep}{0.06in}
\tablecaption{Catalog of Sources in the 70-month \swiftbat\ Survey \label{table_sources}}
\tablewidth{0pt}
\tablehead{
\colhead{Num}& \colhead{BAT Name\tablenotemark{a}}& \colhead{RA\tablenotemark{b}}& \colhead{Dec}& \colhead{S/N}& \colhead{Counterpart Name}& \colhead{Other Name} & \colhead{Ctpt RA\tablenotemark{c}} & \colhead{Ctpt Dec} & \colhead{Flux\tablenotemark{d}} &\colhead{error range\tablenotemark{e}} &\colhead{C\tablenotemark{f}} &\colhead{$\Gamma$\tablenotemark{g}} &\colhead{$\Gamma_{err}$\tablenotemark{e}} &\colhead{$\chi^2_r $} &\colhead{$z$\tablenotemark{h}} & \colhead{Lum\tablenotemark{i}} & \colhead{AS\tablenotemark{j}} &\colhead{Cl\tablenotemark{k}} &\colhead{Type}
}
\startdata
   1 &SWIFT J0001.0$-$0708  &  0.261 & $-$7.123 &   6.10 &2MASX J00004876-0709117     &                                &  0.2032 & $-$7.1532 &  13.03 &   9.05--17.56   &     & 2.17 & 1.75--2.69  &  0.60 &         &      &  & 2 &Galaxy              \\
   2 &SWIFT J0001.6$-$7701  &  0.326 &$-$77.001 &   5.41 &Fairall 1203                &                                &  0.4419 &$-$76.9540 &  10.10 &   6.61--14.08   &     & 2.02 & 1.54--2.60  &  0.70 &  0.0584 &43.92 &  & 4 &Sy1                 \\
   3 &SWIFT J0002.5$+$0323  &  0.664 &    3.332 &   5.10 &NGC 7811                    &                                &  0.6103 &    3.3519 &  11.69 &   7.36--16.64   &     & 1.82 & 1.31--2.39  &  0.50 &  0.0255 &43.24 &  & 4 &Sy1.5               \\
   4 &SWIFT J0003.3$+$2737  &  0.862 &   27.676 &   5.03 &2MASX J00032742+2739173     &                                &  0.8643 &   27.6548 &  13.00 &   8.76--17.70   &     & 1.66 & 1.25--2.12  &  1.00 &       ? &      &  & 2 &Galaxy              \\
   5 &SWIFT J0005.0$+$7021  &  1.011 &   70.327 &   7.20 &2MASX J00040192+7019185     &IGR J00040+7020                 &  1.0082 &   70.3217 &  12.67 &   9.38--16.33   &     & 2.19 & 1.83--2.61  &  0.70 &  0.0960 &44.47 &  & 5 &Sy2                 \\
   6 &SWIFT J0006.2$+$2012  &  1.580 &   20.178 &  10.04 &Mrk 335                     &                                &  1.5813 &   20.2029 &  18.43 &  14.79--22.42   &     & 2.32 & 2.03--2.65  &  0.30 &  0.0258 &43.45 &4 & 4 &Sy1.2               \\
   7 &SWIFT J0009.4$-$0037  &  2.316 & $-$0.575 &   4.99 &2MASX J00091156-0036551     &                                &  2.2982 & $-$0.6152 &   9.26 &   5.31--13.99   &     & 2.14 & 1.52--2.97  &  0.40 &  0.0733 &44.08 &  & 5 &Sy2                 \\
\enddata
\end{deluxetable}
\renewcommand{\arraystretch}{1.0}

\clearpage

\end{document}